\documentstyle[aps,prl]{revtex}
\input psfig.sty

\title{Proton$-$deuteron  elastic scattering above the deuteron breakup}
\author{A. Kievsky, S. Rosati and M. Viviani}
\address{Istituto Nazionale di Fisica Nucleare, Via Buonarroti 2,
 56100 Pisa, Italy}
\address{Dipartimento di Fisica, Via Buonarroti 2, 56100 Pisa, Italy}

\def\x{{\bf x}}
\def\y{{\bf y}}
\def\r{{\bf r}}
\def\a{{\alpha}}
\def\r{{\rho}}
\def\aa{{\alpha\alpha'}}
\def\kk{{kk'}}
\def\tri{{{}^3{\rm H}}}
\def\hel{{{}^3{\rm He}}}
\def\be{\begin{equation}}
\def\ee{\end{equation}}
\def\bea{\begin{eqnarray}}
\def\eea{\end{eqnarray}}
\def\ra{\rightarrow}

\begin{document}

\maketitle
\begin{abstract}
The complex Kohn variational principle and the (correlated)
hyperspherical harmonics method are applied to study the 
proton--deuteron elastic scattering at energies 
above the deuteron breakup threshold. Results for 
the elastic cross section and various elastic
polarization observables have been obtained by taking into
account the long--range effect of the Coulomb interaction
and using a realistic nucleon--nucleon interaction model.
Detailed comparison to the
accurate and abundant elastic proton--deuteron experimental data can now be
performed. 
\end{abstract}


\narrowtext

A number of nucleon--nucleon (NN) potentials are now available
which can be used in a non-relativistic approach to understand
nuclear structure. Those usually referred to as realistic
potentials accurately fit the data base selected by 
the Nijmegen group~\cite{database} with a $\chi^2$ per datum close
to one.  They produce quite reasonable  
values for the binding energies of different light nuclei  
with  small deviations from the experimental values.  
One way of reducing these deviations is to add
three-nucleon (3N) interaction  terms 
determined in a semiphenomenological way. 

When such potentials are used to calculate scattering states,
some observables, such as the elastic $N-d$ differential cross
sections, are well predicted~\cite{report,KRTV96}. However, the
situation is different for a number of  observables which strongly
depend on the nuclear interaction in specific waves. Examples of these
are spin-dependent observables, such as the vector  or the tensor analyzing
powers. For these quantities there are significant differences between
theoretical  estimates and experimental values. This is a strong signal
that  there are deficiencies in the theoretical models adopted. 
Detailed theoretical and experimental investigations are therefore
necessary in order to understand the reasons for this
problem.

In this respect, the study of the $p-d$ scattering
process is of particular relevance since very accurate measurements
exist for a large set of observables and kinematical regimes.
The Faddeev theory has been very successfully applied to the $n-d$
process~\cite{report,ISSU84,CPFG85}  but the extension to the 
$p-d$ case, taking properly into account the long--range Coulomb repulsion,   
presents a number of difficulties which have been the object of
extensive researches~\cite{Alt,Merkuriev}. 
Accurate calculations of $N-d$ scattering below the deuteron
breakup threshold (DBT) have also been performed in the frame of 
the so-called pair--correlated hyperspherical harmonic (PHH) expansion 
technique~\cite{KRV94}. The incorporation of the Coulomb potential
in the PHH approach does not present particular difficulties, and 
$p-d$ scattering observables have been calculated using realistic NN+3N
potentials (see refs. \cite{KRTV96} and \cite{KRV94}).
The technique used is variational and based on the 
use of the Kohn variational principle (KVP).
The extension to  the study of the four--body ($p-\hel$ and $n-\tri$)
zero energy scattering has been given in ref.~\cite{VRK98}.

In ref.~\cite{KRV97} the authors used the complex form of the KVP
to describe $n-d$ and $p-d$
scattering above the DBT, using a  semi-phenomenological
nuclear $s$-wave potential, while the Coulomb  interaction was
included without any partial wave projection.
For the $n-d$ case, the results obtained were in close agreement with
the benchmarks obtained solving the Faddeev equations
in configuration and momentum space~\cite{friar1}.

In the present paper the study of $N-d$ scattering above the DBT
is extended to the case of realistic NN potentials.
Cross sections as well as vector and tensor polarization observables 
for $n-d$ and $p-d$ scattering for  nucleon
incident energies up to $10$ MeV have been calculated
using the NN Argonne AV18 potential~\cite{wiringa}.  These results
are compared to 
the available experimental data~\cite{sagara,exp}.
Some of the $n-d$ results are compared with the values obtained 
by the Bochum-Cracow group~\cite{report}.

The details of the variational approach used by us can be found
in refs.~\cite{KRV94,KRV97,kie97}. The applicability of the
KVP above the DBT when the Coulomb interaction is
taken into account deserves some attention. 
Here a brief discussion of its validity for the description of elastic
scattering is given.
A more general discussion will be reported elsewhere~\cite{KRV99}.
The scattering wave function (w.f.) $\Psi$ is 
written as a sum  of two terms. The first term, $\Psi_C$, describes the
system when the three--nucleons are close to each other. For large
interparticle separations and energies below the 
DBT, it goes to zero, whereas for higher energies it must
reproduce a three outgoing particle state. It
is written as a sum of three Faddeev--like amplitudes
corresponding to even permutations of the particle indices
1, 2, 3. Each amplitude $\Psi_C(\x_i,\y_i)$, where $\x_i,\y_i$ are
the Jacobi coordinates corresponding to the $i$-th permutation, has
total angular momentum $JJ_z$ and total isospin $TT_z$ and it is
decomposed into $N_c$ channels using the $LS$ coupling, namely
\begin{eqnarray}
     \Psi_C(\x_i,\y_i) &=& \sum_\alpha^{N_c} \phi_\alpha(x_i,y_i) 
     {\cal Y}_\alpha (jk,i)  \\
     {\cal Y}_\alpha (jk,i) &=&
     \Bigl\{\bigl[ Y_{\ell_\alpha}(\hat x_i)  Y_{L_\alpha}(\hat y_i) 
     \bigr]_{\Lambda_\alpha} \bigl [ s_\alpha^{jk} s_\alpha^i \bigr ]
     _{S_\alpha}
      \Bigr \}_{J J_z} \; \bigl [ t_\alpha^{jk} t_\alpha^i \bigr ]_{T T_z},
\end{eqnarray}
where $x_i,y_i$ are the moduli of the Jacobi coordinates and
${\cal Y}_\alpha$ is the angular-spin-isospin function for each channel.
The two-dimensional amplitude $\phi_\alpha$ is expanded in terms of the
PHH basis
\begin{equation}
     \phi_\alpha(x_i,y_i) = \rho^{\ell_\alpha+L_\alpha-5/2}f_\alpha(x_i) 
     \left[ \sum_K u^\alpha_K(\rho)\, {}^{(2)}P^{\ell_\alpha,L_\alpha}_K(\phi_i)
     \right] \ ,
\label{eq:PHH}
\end{equation}
where the hyperspherical variables are defined by the relations
$x_i=\rho\cos{\phi}_i$ and $y_i=\rho\sin{\phi}_i$,
$f_\alpha(x_i)$ is a pair correlation function and
${}^{(2)}P^{\ell,L}_K(\phi)$ is a hyperspherical polynomial.

The second term in the variational scattering w.f.  
describes the asymptotic  motion of a deuteron relative to the third
nucleon. It can also be written  as a sum
of three amplitudes in terms of the ingoing and outgoing solutions
of the asymptotic N-d Schroedinger equation.
\begin{equation}
\Omega^+_{LSJ}(\x_i,\y_i) =  \Omega^{in}_{LSJ}(\x_i,\y_i)-
 \sum_{L'S'}{}^J{S}^{SS'}_{LL'}\Omega^{out}_{L'S'J}(\x_i,\y_i)  \ ,
\end{equation}
where ${}^J{S}^{SS'}_{LL'}$ are the elastic $S$--matrix elements.

The three-nucleon scattering w.f. for an incident 
state with relative angular momentum $L$, spin $S$ and total angular momentum
$J$ is
\begin{equation}
\Psi^+_{LSJ}=\sum_{i=1,3}\left[ \Psi_C(\x_i,\y_i)+\Omega^+_{LSJ}(\x_i,\y_i)
             \right] \ ,\label{eq:psip}
\end{equation}
and its complex conjugate is $\Psi^-_{LSJ}$.
A variational estimate of the trial parameters in the w.f. $\Psi^+_{LSJ}$
can be obtained by requiring, in accordance to the complex KVP,
that the functional
\begin{equation}
  [{}^J{S}^{SS'}_{LL'}]= {}^J{S}^{SS'}_{LL'}+i
  \langle\Psi^-_{LSJ}|H-E|\Psi^+_{L'S'J}\rangle \ ,
  \label{eq:kohn}
\end{equation}
be stationary. 

The validity of the KVP above the DBT and with charged particles for the
elastic $S$--matrix elements is briefly discussed below. Let us consider the
w.f. $\overline\Psi^+_{LSJ}$ describing the $p-d$ process 
for an energy $E$, and a trial approximation of it $\Psi^+_{LSJ}$.
Both w.f. can be written in the form 
given in eq.~(\ref{eq:psip}), with the assumption that for the exact one the
sum in eq.~(\ref{eq:PHH}) is not truncated at any level. The 
hyperradial functions and $S$--matrix coefficients entering the w.f. 
$\overline\Psi^+_{LSJ}$ 
will be specified by an overline to distinguish them from the 
corresponding trial quantities.
In the asymptotic region $\rho\ra\infty$, the hyperradial functions 
are superpositions of ingoing and outgoing waves
\bea
     \rho^{\ell_\alpha+L_\alpha}
  u^\a_K(\rho)&\rightarrow& \sum_{\alpha'K'}
  (e^{+i\chi\log2Q\rho})^{KK'}_{\alpha\alpha'}B^{\alpha'}_{K'} e^{-iQ\rho}
   \nonumber \\
  &-& \sum_{\alpha'K'}
   (e^{-i\chi\log2Q\rho})^{KK'}_{\alpha\alpha'}A^{\alpha'}_{K'} e^{iQ\rho}\ ,
    \label{eq:bc}
\eea
where $Q^2=M_NE/\hbar^2$ and the $\chi$--matrix originates from 
the Coulomb potential. Since we are interested in the process
$p+d\ra(p+d)+(p+p+n)$, the boundary conditions to be imposed 
are
\be
   B^\a_K=0\ , \qquad {\rm for\ all\ } K,\a\ .\label{eq:bc2}
\ee

For $\rho\ra\infty$ we can specify four regions, characterized by different 
ranges of values of the hyperangular variables $[\Omega=\phi,\hat x,\hat y]$. 
The $[\Omega_b]$ region is the breakup region where
all the particles are well separated. The region where the particles $j$ and 
$k$ are close each other, while particle $i$ is very far from them, 
is hereafter denoted by $[\Omega_i]$. There are three such regions, 
corresponding to the cases $i=1,2,3$. 
Let us consider the integral
\be\label{eq:kohn1}
  I=\langle
  \overline\Psi^{-}_{LSJ} |(H-E) {\Psi}^{+}_{L'S'J}
   \rangle_R - \langle
  {\Psi}^{-}_{L'S'J}| (H-E) \overline\Psi^{+}_{LSJ}
  \rangle_R\ ,
\ee
where $\langle\rangle_R$ stands for the integration in the
six--dimensional volume with $\rho\le R$ (and
$R\rightarrow\infty$). Only the differential operators present in 
$H$ contribute to $I$. After integrating by parts, the contributions 
come from the hypersurface at $\rho=R$ where
the trial and exact wave functions have reached their
asymptotic behavior.

Let us write $I=I_b+\sum_{i=1}^3 I_i$, where $I_b$ ($I_i$) 
is the contribution
coming from the region $[\Omega_b]$ ($[\Omega_i]$).
In $[\Omega_b]$, the asymptotic
functions $\Omega^+_{LSJ}$ are vanishingly small and $I_b$ reduces to
\be
  I_b \propto \sum_{K\alpha}
   \biggl \{  \bar u^{\a}_{K}(\rho)
    {d \over d \rho} u^{\a}_{K}(\rho) - 
     u^{\a}_{K}(\rho)
     {d \over d \rho} \bar  u^{\a}_{K}(\rho)
   \biggr\}_{R}\ .\label{eq:kohnb1} 
\ee
The above form has been obtained after orthonormalizing the PHH basis
elements at $\rho=\infty$.
Using the asymptotic behavior given in eqs.~(\ref{eq:bc},\ref{eq:bc2})
for both the exact and trial 
hyperradial functions, $I_b\ra 0$ as $R\ra\infty$.
In the three regions $[\Omega_i]$, the breakup part of the w.f. 
can be neglected since it gives contributions which go to zero
as $R^{-3/2}$, therefore
$\sum_{i=1}^3 I_i\propto {}^J\overline{S}^{SS'}_{LL'}
-{}^J{S}^{SS'}_{LL'} $.
Finally, using the fact that $(H-E)\overline\Psi^+_{LSJ}=0$,
it is possible to show that the functional 
$[{}^J{S}^{SS'}_{LL'}]$ differs from ${}^J\overline{S}^{SS'}_{LL'}$
only quadratically in the difference $\epsilon=\overline{\Psi}-\Psi$.

The crucial points of the proof are $i)$ the outgoing
boundary conditions satisfied by $\Psi_C$ and $ii)$ the null contribution
to $I_i$ of the breakup part. The presence of the Coulomb
potential introduces a distortion in the outgoing waves
which, essentially, does not change the main points of the
demonstration of the KVP given for the elastic part of the $S$--matrix
in the n-d case~\cite{nutal}. 

The variation of the functional with respect to the hyperradial 
functions leads to the following set of coupled equations
\bea
  \lefteqn{\sum_{\alpha',k'}
       \Bigl[ A^\aa_\kk (\r ){d^2\over d\r^2}+ B^\aa_\kk (\r ){d\over d\r}
        + C^\aa_\kk (\r )+} && \nonumber \\
     && \qquad {M_N\over\hbar^2} E\; N^\aa_\kk (\r )\Bigr ]
        u^{\alpha'}_{k'}(\r)= D^\lambda_{\alpha k}(\rho) \ .
   \label{eq:siste}
\eea
For each asymptotic state $^{(2S+1)}L_J$  two different inhomogeneous terms
can be constructed corresponding to the
asymptotic $\Omega^\lambda_{LSJ}$ functions with
$\lambda\equiv in,out$. The numerical technique used to solve the above
set of equations imposing outgoing boundary conditions 
at a finite value of the hyperradius $\rho=\rho_0$
is given in ref.\cite{KRV97}. Essentialy, 
the solutions of eq.~(\ref{eq:siste}) for $\rho>\rho_0$ are obtained as
series in $1/\rho$ imposing the outgoing boundary conditions of 
eqs.(\ref{eq:bc},\ref{eq:bc2}).
In the case of $n-d$ scattering such solutions evolve as outgoing Hankel
functions $H^{(1)}(Q\rho)$. In the region $\rho\le\rho_0$
the hyperradial functions have been expanded as
\begin{equation} \label{eq:M}
 \rho^{-5/2}u^\alpha_K(\rho)=\sum_{m=0}^M A^m_{\alpha,K} L^{(5)}_m(z)\exp(-z)
 +A^{M+1}_{\alpha,K} \tilde u_{\alpha,K}(\rho) \ ,
\end{equation}
where $z=\gamma\rho$ and $\gamma$ is a nonlinear parameter.
The functions $L^{(5)}_m(z)$ are Laguerre polynomials.
The parameters $A^m$ and $\gamma$ 
are determined by the variational procedure.
The functions defined above are matched to the outgoing solutions
at $\rho_0$.  The value of the matching radius $\rho_0$ is not critical 
and a value of $\rho_0\approx 100$ fm has been found to be 
satisfactory.

The functions $\tilde u_{\alpha,k}(\rho)$ are the solutions
of eq.~(\ref{eq:siste}) where all the couplings between the differential
equations have been neglected (and applying outgoing boundary
conditions). Their inclusion is necessary since 
the functions $u_K^\alpha(\rho)$ show an oscillatory behavior already
for $\rho>30$ fm. To reproduce such a behavior would require a
rather large value for $M$  in eq.~(\ref{eq:M}). 
However, the inclusion of the terms $\tilde u_{\alpha,K}$ allows
values of $M$ similar to those  needed
for describing $N-d$ scattering below the DBT~\cite{kie97}.

In order to check the convergence properties of the PHH expansion for
energies above the DBT we first solved the same problem treated in
ref.~\cite{KRV97} using the present technique. All phase-shift
and inelasticity parameters were reproduced with the same previous
accuracy, i.e.  with a precision of four figures. 
As compared with ref.~\cite{KRV97}, the dimensions of the matrices
involved  in the eigenvalue problem came out reduced
by one order of magnitude.

Let us start studying $N-d$ scattering above the DBT using the AV18
interaction. The nuclear
elastic $S$--matrix has been calculated up to total angular
momentum states $J=11/2^+$. This includes all partial waves with
relative angular momentum $L \le 4$. Higher partial waves (up to $L=8$)
were included  in the calculation of the observables using the Born
approximation~\cite{KRTV96}.
For each $J^\pi$ state all channels with $\ell_\alpha + L_\alpha \le K_0$
have been included. The number of hyperradial functions
has been increased until convergence was reached. The maximum value
$K_0=6$ was found appropriate to obtain the elastic scattering 
observables within an accuracy of $1\%$. The pattern of convergence in
terms of $K_0$ was studied in ref.~\cite{kie98} for energies below
the DBT and a similar behavior has been observed here.

High quality measurements of $p-d$ scattering have been presented in
ref.~\cite{sagara}. Cross sections 
and proton analyzing powers have been measured up to $E_{lab}=18$ MeV, 
deuteron analyzing powers and tensor analyzing powers 
up to $E_{lab}=9$ MeV ($E_d=18$ MeV).
In fig.~1 our theoretical predictions for these observables are compared
to the data at $E_{lab}=5$ MeV.
In fig.~2 the same set of observables at $E_{lab}=10$ MeV are compared
to the data of ref.~\cite{exp}. 
In addition to the $p-d$
calculations (solid line), the $n-d$ results (dashed line) are also shown
for the sake of comparison. 
A good agreement  between theory and experiment is
observed for the differential cross section. The already known puzzle
has been found again for the vector analyzing powers $A_y$ and
$iT_{11}$ which are underpredicted by about $30\%$. The tensor analyzing
powers are rather well described, with  small underpredictions
at the second minimum in $T_{20}$, the second maximum in $T_{21}$ and the
minimum in $T_{22}$. These differences increase with  energy. 
The origin of these discrepancies can be analyzed in terms of 
phase shift and mixing parameters. For example, in ref.~\cite{KRTV96} 
phase-shift analyses were performed at $E_{lab}=2.5$ MeV and $3.0$ MeV
with the conclusion that small differences in the $P$-wave phase-shifts 
and mixing parameters were responsible for the discrepancy in the  $A_y$
and $iT_{11}$ observables. This problem with the $P$-wave parameters
seems to persist also  at higher energies. The small
discrepancies in the tensor observables could originate from
higher partial waves.  In fact, the tensor observables
are particularly sensitive to phase-shift and mixing parameters for $L\ge 2$.
At $E_{lab}\le 3.0$ MeV, just below the DBT, these parameters
are small due to centrifugal barrier effects, but at the energies considered
here their contribution becomes appreciable. 

Faddeev calculations in momentum space for $n-d$ elastic
scattering at $E_{lab}=5$ and $10$ MeV have been presented in
ref.~\cite{report} for several potential models including the AV18
potential. Our corresponding  results
are in complete agreement with those reference calculations.

In conclusion, $p-d$ elastic cross sections and polarization
observables have been calculated with a realistic interaction
for energies above the DBT  up to $E_{lab}=10$ MeV
and taking into account Coulomb interaction effects.
Accurate calculations of $p-d$ observables and their
comparison with the available experimental data may give 
stringent tests of the existing models of NN and 3N interactions. 
The extension of the present technique to higher energies and to the
breakup cross sections will be the subject of a forthcoming
paper.

\begin{figure}[t]
\caption{
Differential cross section $d\sigma/d\Omega$, proton analyzing power
$A_y$, deuteron analyzing power $iT_{11}$ and 
tensor analyzing powers $T_{20}$, $T_{21}$ and $T_{22}$
calculated at $E_{lab}=5$ MeV and compared with the data
of ref.~\protect\cite{sagara} (circles with error bars). The solid
(dashed) lines are the $p-d$ ($n-d$) results. 
}
\end{figure}

\begin{figure}[t]
\caption{
As in fig.~1, but for $E_{lab}=10$ MeV. The data reported here are
from ref.\protect\cite{exp}. 
}
\end{figure}

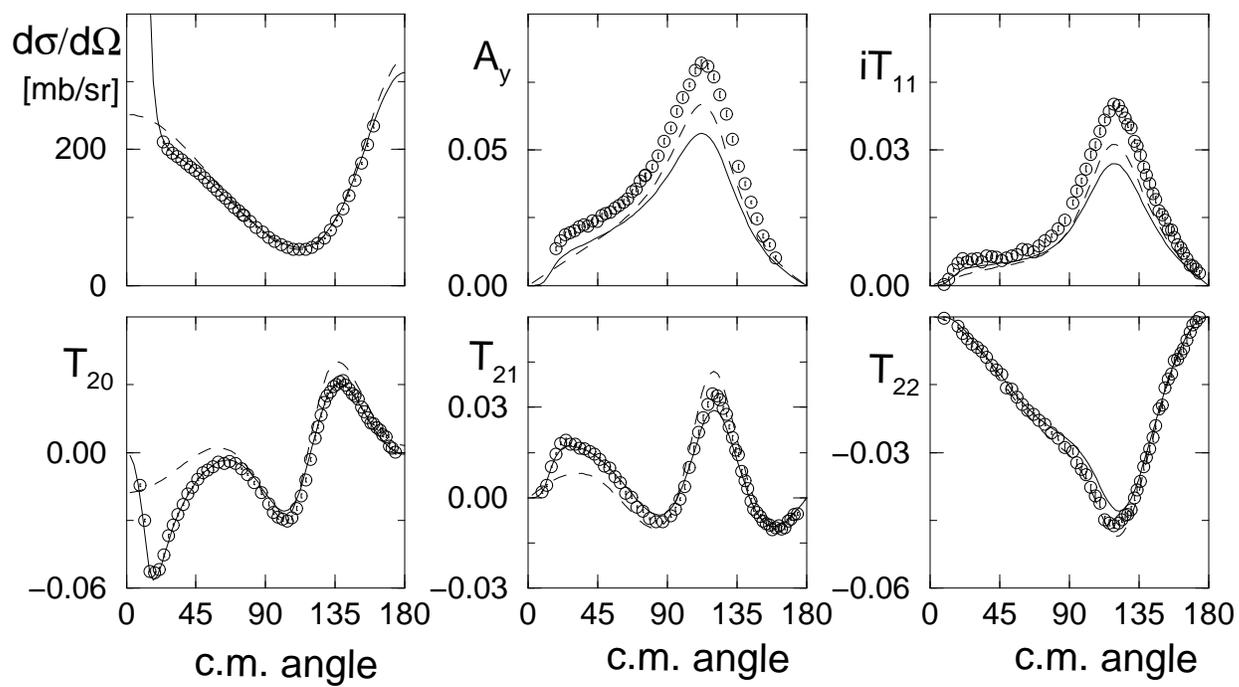
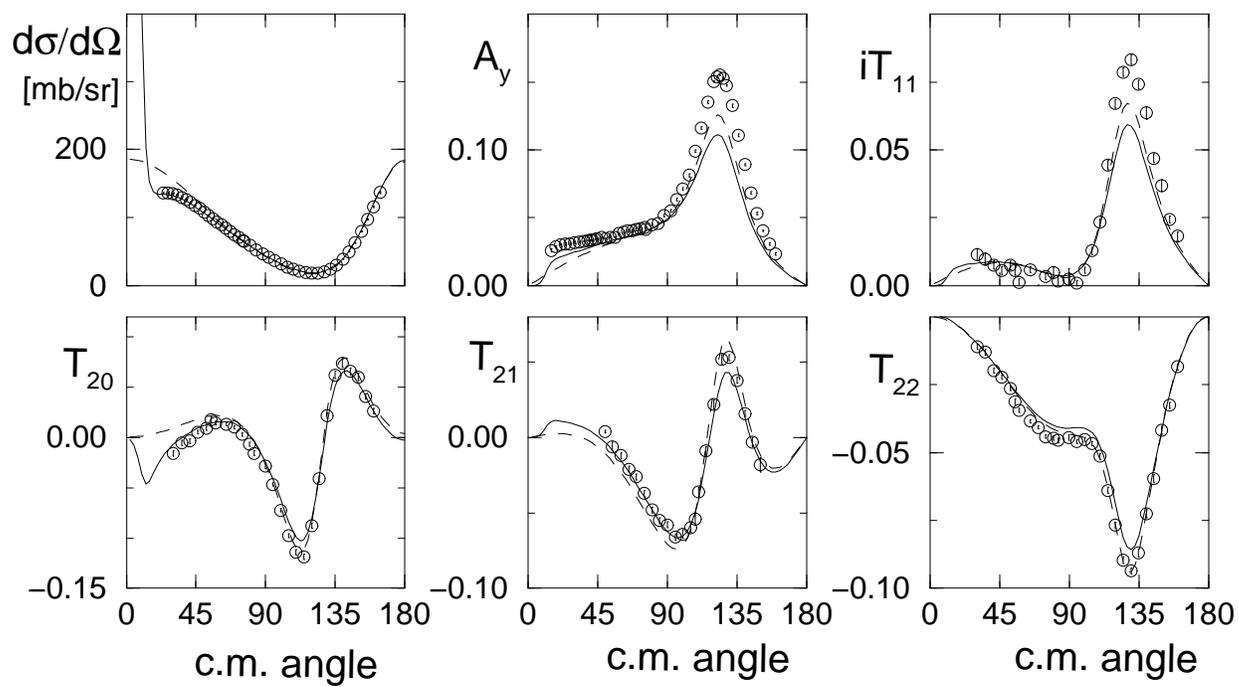
\end{document}